\documentstyle[aps,pra,eqsecnum,floats]{revtex}
  
\def\Z{\angle \!\!\!{\rm Z}}  
  
\def\C{\rm l\!\!\!C\,}  
  
\def\uma{\rm 1\!\!\hskip 1 pt l}  
\tighten
\draft
\begin{document}

\preprint{} 
\title{ An Algebraic Construction of Generalized Coherent 
States for Shape-Invariant Potentials}

\author{A.~N.~F.       Aleixo\thanks{Electronic    address:       {\tt
armando@if.ufrj.br}}} \address{Instituto de F\'{\i}sica,  Universidade
Federal do   Rio     de  Janeiro,   RJ   -    Brazil}    \author{A.~B.
Balantekin\thanks{             Electronic            address:     {\tt
baha@nucth.physics.wisc.edu}}}    \address{Department   of    Physics,
University of Wisconsin, Madison, Wisconsin 53706 USA  
\vskip 1 true cm}

\date{\today} 
\maketitle
\begin{abstract}

\vskip 0.3 true cm

  Generalized coherent states for shape invariant potentials are constructed 
  using an algebraic approach based on supersymmetric quantum mechanics. We show
  this generalized formalism is able to: a) supply the essential requirements 
  necessary to establish a connection between classical and quantum 
  formulations of a given system (continuity of labeling, resolution of unity, 
  temporal stability, and action identity); b) reproduce results already 
  known for shape-invariant systems, like harmonic oscillator, double anharmonic, 
  P\"oschl-Teller and self-similar potentials and; c) point to a formalism 
  that provides an unified description of the different kind of coherent 
  states for quantum systems.
\end{abstract}



\newpage

\section{Introduction}

Coherent states were first introduced by Schr\"odinger \cite{r1}, who was interested 
in finding quantum-mechanical states which provide a close connection between quantum 
and classical formulations of a given physical system. Based on the Heisenberg-Weyl 
group and applied specifically to the harmonic oscillator system, the original coherent 
state introduced by Schr\"odinger has been extended to a large number of Lie
groups with square integrable representations \cite{r2,r3}. Today these 
extensions represent many applications in a number of fields of quantum theory, 
and especially in quantum optics and radiophysics. In particular they are 
used as bases of coherent state path integrals \cite{r4} or dynamical 
wavepackets for describing the quantum systems in semi-classical approximations
\cite{r5}. There are different definitions 
of coherent states. The first one, often called Barut-Girardello coherent states \cite{r5a}, 
assumes the coherent states are eigenstates with complex eigenvalues of an annihilation 
group operator. The second definition, often called Perelomov coherent states \cite{r5b},  
assumes the existence of an unitary $z$-displacement operator whose action on the 
ground state of the system gives the coherent state parameterized by $z$, with $z\in\C$.
The last definition, based on the Heisenberg uncertainty relation, often called {\it 
intelligent} 
coherent states \cite{r5c}, assumes that the coherent state gives the 
minimum-uncertainty value \ $\Delta x\,\Delta p = {\hbar\over 2}\,,$ and maintains this 
relation in time because its temporal stability. These three different definitions are 
equivalent only in the special case of the Heisenberg-Weyl group, the dynamical 
symmetry group of the harmonic oscillator. 

The extension of coherent states for systems other than harmonic oscillator 
has attracted much attention for the past several years \cite{r6,r7,r8,r9,r10,r11,r11a}. 
There are a large number of different approaches to this problem and the one 
to be presented here is based on the supersymmetric quantum mechanics.
Supersymmetric quantum mechanics \cite{r12} deals with pairs of
Hamiltonians which have the same energy spectra, but different
eigenstates. A number of such pairs of Hamiltonians share an
integrability condition called shape invariance \cite{r13}. Although
not all exactly-solvable problems are shape-invariant \cite{r14},
shape invariance, especially in its algebraic formulation
\cite{r15,r16}, is a powerful technique to study exactly-solvable 
systems. 

Supersymmetric quantum mechanics is generally studied in the context
of one-dimensional systems. The partner Hamiltonians
\begin{equation}
\label{eqh12}
\hat H_{_1} = -{\hbar^2\over 2m}{d^2\over dx^2}+V_-(x) = \hbar\Omega\,\hat A^\dagger\hat A  
\qquad\qquad {\rm and}\qquad\qquad
\hat H_{_2} = -{\hbar^2\over 2m}{d^2\over dx^2}+V_+(x) = \hbar\Omega\,\hat A\hat A^\dagger
\end{equation}
are most readily written in terms of one-dimensional operators
\begin{equation}
\label{eqaa+}
\hat A \equiv {1\over \sqrt{\hbar\Omega}}\left(W(x) + \frac{i}{\sqrt{2m}}\hat p \right)
\qquad\qquad {\rm and}\qquad\qquad
\hat A^\dagger \equiv {1\over\sqrt{\hbar\Omega}}\left(W(x) - \frac{i}{\sqrt{2m}}\hat p\right)
\end{equation}
where $\hbar\Omega$ is a constant energy scale factor, introduced to permit working with 
dimensionless quantities, and $W(x)$ is the superpotential which is related to the 
potentials $V_\pm(x)$ via
\begin{equation}
V_\pm(x) = W^2(x)\pm{\hbar\over \sqrt{2m}}{dW(x)\over dx}\,.
\label{eqvpm}
\end{equation}
 
In earlier works, by using an algebraic approach, we introduced coherent states for 
self-similar potentials \cite{r10}, a class of shape-invariant systems, and presented a 
possible generalization of these coherent-states and its relation with the Ramanujan's 
integrals \cite{r11}. In the present paper we extend this generalized formalism to all 
shape-invariant systems and show that the generalized coherent-states then obtained 
satisfy the essential requirements necessary to provide the basic principles 
\cite{r17} embodied in Schr\"odinger's original idea.
This paper is organized as follows. In Section II we present the 
algebraic formulation to shape invariance and introduce the 
fundamental principles of our generalized coherent states and its basic properties; 
in Section III we apply our general formalism to shape-invariant systems classified
using the factorization method introduced by Infeld and Hull \cite{r18} and work out
some possible examples of coherent states for these systems. Finally, brief 
remarks close the paper in Section IV.


\section{Generalized Coherent States for Shape-Invariant Systems}

The Hamiltonian $\hat H_{_1}$ of Eq.~(\ref{eqh12}) is called
shape-invariant if the condition
\begin{equation}
\hat A(a_{_1}) \hat A^\dagger(a_{_1}) =\hat A^\dagger (a_{_2})  \hat A(a_{_2}) +
R(a_{_1}) \,,
\label{eqsi}
\end{equation}
is satisfied \cite{r13}. In this equation $a_{_1}$ and $a_{_2}$ represent
parameters of the Hamiltonian. The parameter $a_{_2}$ is a function of
$a_{_1}$ and the remainder $R(a_{_1})$ is independent of the dynamical
variables such as position and momentum. As it is written the
condition of Eq.~(\ref{eqsi}) does not require the Hamiltonian to be
one-dimensional, and one does not need to choose the ansatz of
Eq.~(\ref{eqaa+}). In the cases studied so far the parameters $a_{_1}$
and $a_{_2}$ are either related by a translation \cite{r15,r19} or a
scaling \cite{r10,r11,r20}. Introducing the similarity 
transformation that replace $a_{_1}$ with $a_{_2}$ in a given operator 
\ $\hat T(a_{_1})\, \hat O(a_{_1})\, \hat T^\dagger(a_{_1}) = \hat O(a_{_2})$ \ 
and the operators \ $\hat B_+ =  \hat A^\dagger(a_{_1})\hat T(a_{_1})$ \ and \ 
$\hat B_- =\hat B_+^\dagger =  \hat T^\dagger(a_{_1})\hat A(a_{_1})\,,$
the Hamiltonians of Eq.~(\ref{eqh12}) take the forms \ $\hat H\equiv \hat H_{_1} 
=\hbar\Omega\,\hat B_+\hat B_-$ \ 
and \ $\hat H_{_2} = \hbar\Omega\,\hat T\hat B_-\hat B_+\hat T^\dagger\,.$ \ 
As shown in  \cite{r15}, with Eq.~(\ref{eqsi}) one can also easily prove the commutation
relation \ $[\hat B_-,\hat B_+] =  \hat T^\dagger(a_{_1})R(a_{_1})\hat T(a_{_1})  \equiv
R(a_{_0})\,,$ \ where we used the identity \ $R(a_n) = {\hat T}(a_{_1})\,R(a_{n-1})
\,{\hat T}^\dagger (a_{_1})\,,$ \ 
valid for any $n\in\Z$. This commutation relation suggests that $\hat{B}_-$ and $\hat{B}_+$ 
are the appropriate creation and annihilation operators for the spectra of 
the shape-invariant potentials provided that their non-commutativity with 
$R(a_{_1})$ is taken into account. The additional relations 
\begin{equation}
\label{eqrbpm}
R(a_n) \hat{B}_+ = \hat{B}_+ R(a_{n-1}) \qquad\qquad {\rm and}\qquad\qquad
R(a_n) \hat{B}_- = \hat{B}_- R(a_{n+1})\,,
\end{equation}
readily follow from these results. Considering that the ground state of the Hamiltonian 
$\hat H$ satisfies the condition
\begin{equation}
\hat A\,\vert \Psi_{_0}\rangle = 0 = \hat B_-\,\vert \Psi_{_0}\rangle\,,
\label{eqaps0}
\end{equation}
using the relations above it is possible to find the $n$-th excited state of $\hat H$ 
\begin{equation}
\label{eqh12n}
\hat H\vert \Psi_n\rangle \equiv \hbar\Omega\left(\hat B_+\hat B_-\right)\vert \Psi_n\rangle = 
\hbar\Omega\,{e}_{n}\,\vert \Psi_n\rangle 
\quad\quad {\rm and}\quad\quad
\hat B_-\hat B_+\vert \Psi_n\rangle = 
\left\{{e}_n+R(a_{_0})\right\}\vert \Psi_n\rangle\,.
\end{equation}
where these eigenstates can be written in a normalized form as 
\begin{equation}
\label{eqpsn}
\vert \Psi_n \rangle = \frac{1}{\sqrt{R(a_{_1})+ R(a_{_2})+\cdots + 
R(a_n)}} \hat{B}_+ \cdots \frac{1}{\sqrt{R(a_{_1})+ 
R(a_{_2})}} \hat{B}_+  \frac{1}{\sqrt{R(a_{_1})}}\hat{B}_+
\vert \Psi_{_0} \rangle 
\end{equation}
with the eigenvalues \ $E_n = \hbar\Omega\,e_n$, \ being
\begin{equation}
e_n =  \sum_{k=1}^n R(a_k)\,.
\label{eqen}
\end{equation}

As mentioned in the introduction, a possible way to define a coherent state is to find a 
quantum state annihilated by the lowering operator. Annihilation-operator coherent states 
for shape-invariant potentials were introduced in \cite{r7,r10}. Here we follow the notation 
of \cite{r10}. Our first step is to introduce the necessary tools to be used in this
construction. After we obtain the coherent-state we must verify if this state 
satisfies the set of four essential requirements, introduced and
discussed in \cite{r17}, necessary for a close connection between classical and 
quantum formulations of a given system: a) label continuity; b) overcompleteness or resolution 
of unity; c) temporal stability and; d) action identity. Indeed the first two requirements are 
standard and rely on the algebraic structure behind the system in question, while
the last two are more general and relate to the classical-quantum connection question.


\subsection{Construction}

To remove the energy scale we rewrite the shape-invariant Hamiltonian as
\begin{equation}
\hat H = \hbar\Omega\,\hat {\cal H}\,, \qquad {\rm with}\qquad \hat {\cal H} = \hat B_+\hat B_-\,.
\label{eqhdims}
\end{equation}
The operator $\hat B_-$ does not have a left inverse in the
Hilbert space of the eigenstates of the Hamiltonian $\hat H$. However, 
a right inverse for $\hat B_-$ \ $(\hat B_-\hat B_-^{-1} = \hat 1)$, \ can be defined. 
Similarly the inverse of $\hat {\cal H}$ does not exist, but \ $\hat {\cal H}^{-1}\hat B_+ 
= \hat B_-^{-1}$ \ 
does. Therefore, if we define the Hermitian conjugate operators \ $\hat Q = \hat B_-\hat 
{\cal H}^{-1/2}\,,$ \ \ 
and \ \ $\hat Q^\dagger = \hat {\cal H}^{-1/2}\hat B_+$ \ \ 
we can easily show that \ $\hat B_-^{-1} = \hat {\cal H}^{-1/2}\hat Q^\dagger$ \ 
and the normalized form of the $n$-th excited state of $\hat H$, 
given by (\ref{eqpsn}), can be rewritten as \ $\vert \Psi_n \rangle = (\hat Q^\dagger )^n\,
\vert \Psi_{_0} \rangle$.
Then, taking into account Eqs.~(\ref{eqh12n}), (\ref{eqen}) and these two
last relations, we can prove that
\begin{equation}
\hat B_-^{-n}\vert \Psi_{_0} \rangle = C_n\,\vert \Psi_n \rangle\,,\qquad {\rm where}\qquad
C_n = \left\{\prod_{k=0}^{n-1}\left(e_n-e_k\right)\right\}^{-1/2} = 
\left\{\prod_{k=1}^{n}\left[\,\sum_{s=k}^{n}R(a_s)\right]\right\}^{-1/2}\,,
\label{eqcoefn}
\end{equation}
since \ $e_{_0} = 0\,.$ \ After these preliminary considerations we are ready
to define our generalized expression for the coherent state of shape-invariant
systems as
\begin{equation}
\vert z;a_j\rangle = \sum_{n=0}^\infty\left\{z\,{\cal Z}_j\,
\hat B_-^{-1}\right\}^n\,\vert \Psi_{_0}\rangle\,,\qquad\qquad z,\; {\cal  Z}_j\;\in\;{\C}\,,
\label{eqzf}
\end{equation}
where we used the shorthand notation 
${\cal Z}_j \equiv {\cal Z}(a_j)\equiv {\cal Z}(a_{_1},a_{_2},a_{_3},\dots)$ 
for an arbitrary functional of the potential parameters, introduced to establish 
a more general approach. As one will see in the applications below, for harmonic oscillator
system, the presence of the functional ${\cal Z}_j$ introduces only a constant scale 
factor in the complex expansion variable $z$ that can be absorbed by a redefinition of 
this constant and, thus, we get back to the standard results for this system.  
Formally the definition (\ref{eqzf}) can be expressed as
\begin{equation}
\vert z;a_j\rangle = \left[{1\over 1-z{\cal  Z}_j\,\hat B_-^{-1}}\right]
\,\vert \Psi_{_0}\rangle\,.
\label{eqzf1}
\end{equation}
Using the relation (\ref{eqrbpm}) we can prove this coherent state 
is eigenstate of the operator $\hat{B}_-$ since
\begin{equation}
\hat{B}_-\vert z;a_j\rangle = z\,{\cal  Z}_{j-1}\,\vert z;a_j\rangle\,.
\label{eqbmz}
\end{equation}
This state also satisfies the additional condition
\begin{equation}
\left\{\hat{B}_- - z{\cal  Z}_{j-1}\right\}{\partial\over\partial z}\,
\vert z;a_j\rangle = {\cal  Z}_{j-1}\,\vert z;a_j\rangle\,,
\label{eqbzadi}
\end{equation}
where \ ${\cal  Z}_{j-1}= \hat T^\dagger(a_{_1})\,{\cal  Z}_j\,\hat T(a_{_1})\,.$
An important observation is that the coherent state definition (\ref{eqzf}) satisfies
the continuity of labeling requirements since the transformation of the variables
\ $(z,a_j)\rightarrow (z^\prime,a_{j^\prime})$ \ leads to the transformation of the states
\ $\vert z;a_j\rangle\rightarrow \vert z^\prime;a_{j^\prime}\rangle\,.$
This is the first standard property required for coherent states. The other three 
we take up in the next subsections. Together with the resolution of unity, the 
continuity of labeling represents the minimal condition to be satisfied for a
set of coherent states to be represented by a Lie algebraic group. 


\subsection{Normalization}

At this stage we can use the action of the $\hat B_-^{-1}$ operator on the Hilbert space
of the eigenstates \ $\left\{\vert \Psi_n \rangle, \; n = 0,1,2,\dots\right\},$ \ 
 and (\ref{eqrbpm}) to get the generalized Glauber's form \cite{r21} of the coherent state 
$\vert z;a_j\rangle$ based in its expansion in the eigenstates of the Hamiltonian $\hat H$:
\begin{equation}
\vert z;a_r\rangle = {\cal N}\left(\vert z\vert^2;a_r\right)\sum_{n=0}^\infty {z^n\over 
h_n(a_r)}\,
\vert \Psi_n\rangle\,,
\label{eqzfn}
\end{equation}
where we used the shorthand notation \ 
$(a_r) \equiv \left[R(a_{_1}),R(a_{_2}),\dots,R(a_n);a_j,a_{j+1},\dots,a_{j+n-1}\right]$ 
\ for the expansion coefficients, which are given by \ $h_{_0}(a_r) = 1$ \ and
\begin{equation}
h_n\left(a_r\right) = \sqrt{\prod_{k=1}^{n}\left[\,\sum_{s=k}^{n}R(a_s)\right]}
\Biggl/\;\prod_{k=0}^{n-1}{\cal  Z}_{j+k}
\,,\qquad {\rm for}\qquad n\ge 1
\label{eqhnc}
\end{equation}
with \ ${\cal  Z}_{j+k} = \{\hat T(a_{_1})\}^k\, {\cal  Z}_j\,\{\hat T^\dagger(a_{_1})\}^k\,,$
as well for the real normalization factor
\begin{equation}
{\cal N}\left(x;a_r\right) = 1\Biggl/\sqrt{\sum_{n=0}^\infty {x^n\over \vert h_n(a_r)\vert^2}}\,.
\label{eqnorm}
\end{equation}
At this point we observe that the transformation properties between the potential 
parameters $a_n$, imposed by shape invariance, constrains the freedom to define 
${\cal Z}_j$. Besides that, when we consider the relation 
(\ref{eqhnc}), this potential parameter dependence in ${\cal  Z}_j$ shows 
strong influence in the final expression of the expansion coefficient 
$h_n\left(a_r\right)\,.$ Another thing to observe about ${\cal  Z}_j$ is 
its importance in the determination of the radius of convergence in the series 
defining ${\cal N}\left(\vert z\vert^2;a_r\right)$ since this radius is
given by \ ${\cal R} = \limsup_{n\rightarrow+\infty} 
\sqrt[n]{\vert h_n\left(a_r\right)\vert^2}\,.$

It should be noted that this normalized coherent state has an 
$\hat B_-$ operator eigenvalue different from the unnormalized one since the 
potential parameters in the normalization factor are changed by the action of
that operator. Indeed we can prove that in this case Eq.~(\ref{eqbmz}) must 
assume the form
\begin{equation}
\hat{B}_-\vert z;a_{_j}\rangle = z\,{\cal  Z}_{_{j-1}}
\left[{{\cal N}\left(a_{r-1};\vert z\vert^2\right)\over
{\cal N}\left(a_r;\vert z\vert^2\right)}\right]\vert z;a_{_j}\rangle\,,
\label{eqbmzn}
\end{equation}
where \ ${\cal N}\left(a_{r-1};\vert z\vert^2\right)= 
\hat T^\dagger(a_{_1})\,{\cal N}\left(a_{r};\vert z\vert^2\right)\,\hat T(a_{_1})\,.$ \
Although they are normalized, the coherent states $\vert z;a_{_r}\rangle$ are not orthogonal 
to each other since
\begin{equation}
\langle z^\prime;a_r\vert z;a_r\rangle = {{\cal N}\left(a_r;\vert z^\prime\vert^2\right)\,
{\cal N}\left(a_r;\vert z\vert^2\right)\over {\cal N}^2\left(a_r;zz^{\prime *}\right)}\,.
\label{eqzpz}
\end{equation}
So we conclude that they form an over-complete linearly dependent set.


\subsection{Overcompleteness}

Now we can investigate the overcompleteness or resolution of unity property of the generalized 
coherent states introduced by equation (\ref{eqzf}). To this end we assume the existence of a 
positive-definite weight function $w\left(\vert z\vert^2;a_r\right)$ so that an integral over 
the complex plane exists and gives the result
\begin{equation}
\int_{\;\C} d^2z\; \vert z;a_r\rangle\langle z;a_r\vert\; w\left(\vert z\vert^2;a_r\right) = 
{\hat {\uma}}_{_{\cal H}}\,,
\label{eqcomp}
\end{equation}
where ${\hat {\uma}}_{_{\cal H}}$ is the identity operator in the Hilbert space of 
the $\hat H$-eigenstates. Inserting Eq.~(\ref{eqzfn}) into Eq.~(\ref{eqcomp}) the resolution 
of unity can be expressed by
\begin{equation}
\int_{\;\C} d^2z\; {\cal N}^2\left(\vert z\vert^2;a_r\right)\sum_{m,n=0}^\infty {z^{*m}z^n
\over h_m^*(a_r)\,h_n(a_r)}\,
\vert \Psi_n\rangle\langle \Psi_m\vert\; w\left(\vert z\vert^2;a_r\right) = 
{\hat {\uma}}_{_{\cal H}}\,.
\label{eqcomp1}
\end{equation}
At this point we can use the orthonormality of the eigenstates $\vert \Psi_n\rangle$ 
to show that the diagonal matrix elements of Eq.~(\ref{eqcomp1}) can be written as
\begin{equation}
\int_{\;\C} d^2z\; {\cal N}^2\left(\vert z\vert^2;a_r\right)\,(z^*z)^n\,
w\left(\vert z\vert^2;a_r\right) = \vert h_n(a_r)\vert^2\,.
\label{eqcomp2}
\end{equation}
Therefore, assuming the polar coordinate representation \ $z \equiv re^{i\phi}$ \ 
of complex numbers we must have \ $d^2z = r\,dr\,d\phi$ \ and using the result
\begin{equation}
{1\over 2\pi}\int_0^{2\pi}d\phi\;e^{i(n-m)\phi} = \delta_{n,m}
\label{eqdnm}
\end{equation}
we conclude that to get a resolution of unity we must require
\begin{equation}
\int_0^\infty d\rho\, \rho^n\, {\cal W}\left(\rho;a_r\right)  = \vert h_n(a_r)\vert^2\,,
\quad\qquad {\rm where}\quad\qquad 
{\cal W}\left(\rho;a_r\right)=\pi\,{\cal N}^2\left(\rho;a_r\right)\,w\left(\rho;a_r\right)\,,
\label{eqdnnr}
\end{equation}
and $\rho$ stands for $r^2\,.$
In other words, equation (\ref{eqdnnr}) provides the set of moments $\{\rho_n\}$ 
of the distribution function \ ${\cal W}(\rho;a_r)$, \ since we assume all moments 
exist and have finite values. Therefore, as pointed in Ref.~\cite{r22}, the 
problem of finding a suitable measure \ $w(\rho;a_r)$ \ reduces to a moment distribution problem. 
After this point there are several possible ways to get the measure \ $w(\rho;a_r)$. \
We can choose a possible form of \ $w(\rho;a_r)$ \ by using the result of a known integral. 
Another possibility is to use a transformation procedure, like 
Mellin \cite{r23} or Fourier, to determine the form of the measure \ $w(\rho;a_r)$. \ 
For example, in the Fourier transformation case we can multiply Eq.~(\ref{eqdnnr}) by 
the sum factor \ $\sum_{n=0}^\infty (i\xi)^n/n!$ and use the series expansion of the 
exponential function to obtain
\begin{equation}
\int_0^\infty d\rho\, {\cal W}\left(\rho;a_r\right)\, e^{i\rho\xi} = \Phi(\xi;a_r)
 = \sum_{n=0}^\infty {\vert h_n(a_r)\vert^2\,(i\xi)^n/n!}\,.
\label{eqPhi}
\end{equation}
Thus, taking the inverse Fourier transformation of Eq.~(\ref{eqPhi}) we can show that
\begin{equation}
{\cal W}(\rho;a_r) = {1\over 2\pi}\int_{-\infty}^\infty d\xi\;\Phi(\xi;a_r)\,e^{-i\rho\xi}\,.
\label{eqwrho1}
\end{equation}
In the applications of the next section we will use several different procedures  
to get the resolution of unity. To conclude this part, we note that explicit computation of 
the weight function $w\left(\vert z\vert^2;a_r\right)$ requires the knowledge of the 
spectrum of the quantum mechanical system under consideration and the form of the
functional ${\cal  Z}_j$.


\subsection{Temporal Stability}

Let us now investigate the dynamical evolution of the generalized coherent state 
(\ref{eqzfn}). To do that we must remember that the time evolution of this generalized 
coherent state can be obtained by
\begin{equation}
\vert z;a_r\rangle_{_t} = \hat U(t,0)\vert z;a_r\rangle_{_o}
\label{eqchnt}
\end{equation}
where the time evolution operator fulfills the differential equation 
\begin{equation}
i\hbar\,{\partial \hat U(t,0)\over \partial t} = \hat H\,\hat U(t,0)\,,
\label{eqdut}
\end{equation}
with the initial condition \ $\hat U(0,0) = {\hat {\uma}}_{_{\cal H}}\,.$ \ Thus,
\begin{equation}
\vert z;a_r\rangle_{_t} = \exp{\left(-i{\hat H}t/\hbar\right)}\vert z;a_r\rangle_{_o}\,.
\label{eqchnt1}
\end{equation}
At this point if we consider the expansion (\ref{eqzfn}), the results of Eqs.~(\ref{eqh12n}) 
and the commutation between any function of the potential parameters $a$ and the 
Hamiltonian $\hat H$ in the equation (\ref{eqchnt1}) we obtain 
\begin{equation}
\vert z;a_r\rangle_{_t} = {\cal N}\left(\vert z\vert^2;a_r\right)\sum_{n=0}^\infty 
{z^n\over h_n(a_r)}\,
e^{-i\Omega e_n t}\,\vert \Psi_n\rangle\,.
\label{eqzfnt}
\end{equation}
To establish the temporal stability of this coherent state we utilize the freedom in the choice 
of the functional ${\cal  Z}(a_j)$ to redefine it as \ $\bar {\cal  Z}(a_j) = 
{\cal  Z}(a_j)\,e^{-i\alpha R(a_{_1})}$ \ where $\alpha$ is a real constant. 
This redefinition implies \ $\bar h_n(a_r) = h_n(a_r)\,e^{i\alpha e_n}\,,$
where $e_n$ is given by (\ref{eqen}) and $h_n(a_r)$ still given by equation (\ref{eqhnc}).
Therefore we can write the coherent state $\vert z;a_r\rangle$ as
\begin{equation}
\vert z;a_r\rangle\Longrightarrow\vert z,\alpha;a_r\rangle = 
{\cal N}\left(\vert z\vert^2;a_r\right)\sum_{n=0}^\infty {z^n\over h_n(a_r)}\,
e^{-i\alpha e_n}\,\vert \Psi_n\rangle\,,
\label{eqzfnred}
\end{equation} 
and its time-evolved form as
\begin{equation}
\vert z,\alpha;a_r\rangle_{_t} = {\cal N}\left(\vert z\vert^2;a_r\right)\sum_{n=0}^\infty 
{z^n\over h_n(a_r)}\,
e^{-i(\alpha+\Omega t)e_n}\,\vert \Psi_n\rangle\equiv \vert z,\alpha+\Omega t;a_r\rangle\,,
\label{eqzfntred}
\end{equation}
illustrating the fact that the time evolution of any such generalized coherent state remains
within the family of generalized coherent states. In other words, the generalized coherent states 
\ $\vert z,\alpha;a_r\rangle$ \ show temporal stability under $\hat H$. \ 
To conclude this part, note that the polar coordinates representation of the redefined
complex functional $\bar {\cal  Z}(a_j)$ imply that in the coherent state time evolution
its real modulus remains constant while its complex phase increases linearly. These 
properties are similar to the classical behaviour of canonical action-angle variables.


\subsection{Action Identity}

The last property to be satisfied for the coherent state \ $\vert z;a_r\rangle$ \
is the action identity. To verify this identity we take the conjugate of Eq.~(\ref{eqbmz}) 
and use the definition of the operator $\hat B_+$ to get
\begin{equation}
\langle z;a_r\vert\hat{B}_+ = \langle z;a_r\vert z^*\,{\cal  Z}_{j-1}^*\,.
\label{eqkmz}
\end{equation}
Now with this result, Eq.~(\ref{eqbmz}) and the expression of the Hamiltonian $\hat H$ we 
can calculate 
the expectation value
\begin{equation}
\langle\hat H\rangle = {\langle z;a_r\vert\hat H\vert z;a_r\rangle\over
\langle z;a_r\vert z;a_r\rangle} = \hbar\Omega\;
{\langle z;a_r\vert\hat B_+\hat B_-\vert z;a_r\rangle\over
\langle z;a_r\vert z;a_r\rangle} = 
\hbar\Omega\,\vert z\,{\cal  Z}_{j-1}\vert^2\,.
\label{eqacid}
\end{equation}
Using this result we can define a canonical action variable \ 
$J = \hbar\beta_j^*\beta_j\,,$ \ with \ $\beta_j = z{\cal  Z}_{j-1}\,,$
such as \ $\langle\hat H\rangle = \nu J\,,$ \ so that \ 
$\dot\nu = \partial\langle\hat H\rangle/\partial J=\Omega\;\Longrightarrow\; 
\nu = \Omega t+\alpha\,,$ \ as required for a couple of canonical conjugate action-angle 
variables. 
Note that the normalized form (\ref{eqzfn}) of the coherent state 
$\vert z;a_r\rangle$ requests the definition \ 
$\beta_j = z\,{\cal  Z}_{j-1}
{\cal N}\left(\vert z\vert^2;a_{r-1}\right)\bigl/
{\cal N}\left(\vert z\vert^2;a_r\right)\,.$

With these properties we showed that the generalized coherent state 
 \ $\vert z;a_r\rangle$ \ 
satisfies the set of basic requirements we enumerated.


\section{Some Examples of Generalized Coherent State Systems}

Using the definition presented in the previous section we now illustrate the
concept of generalized coherent states for shape invariant systems using 
some known shape invariant potential systems. 
As in reference \cite{r7}, for these applications we follow the 
classification based on the factorization method introduced by Infeld and Hull 
\cite{r18} in which six possibles types of shape-invariant systems are grouped when
its potential parameters are related by a translation. We also study the case of self-similar
potential system as an example of shape invariant potential with potential parameters 
related by scaling. 


\subsection{Types (C) and (D) shape-invariant systems}

We begin with these systems because they are the simplest cases among the shape-invariant 
potential systems. The partner potentials $V_\pm(x)$ for these systems are obtained with
the superpotentials 
\begin{equation}
W_{_{\rm C}}(x,a_{_1}) = \sqrt{\hbar\Omega}\left({a_{_1}+\delta\over x} + {\beta\over 2}x\right)
\,,\qquad {\rm and}\qquad
W_{_{\rm D}}(x,a_{_1}) = \sqrt{\hbar\Omega}\left(\beta x + \delta\right)\,,
\label{eqwcd}
\end{equation}
where $\beta$ and $\delta$ are real constants, while the remainders in the shape-invariant 
condition (\ref{eqsi}) are given by \cite{r12}
\begin{equation}
R_{_{\rm C}}(a_n) = \beta\left(a_n-a_{n+1}+\sqrt{\hbar\over 2m\Omega}\,\right)\,,
\qquad {\rm and}\qquad
R_{_{\rm D}}(a_n) = \sqrt{\hbar\over 2m\Omega}\left(a_n+a_{n+1}\right)\,.
\label{eqwcdr}
\end{equation}
Taking into account that the parameters for these potentials are related by
\begin{equation}
\cases{a_{n+1} = a_n-\sqrt{\hbar/(2m\Omega)}\,,\qquad{\rm for \ \ (C)}\,,\cr
\qquad\qquad\qquad\qquad\qquad\qquad\qquad\qquad\qquad\qquad\qquad\qquad(\forall n\in \Z)\,,\cr
a_{_1} = a_{_2} = \dots = a_n = \beta\,,\qquad\;\,{\rm for \ \ (D)}\,,\cr}
\label{eqancd}
\end{equation}
we conclude that for both shape-invariant systems the remainders (\ref{eqwcdr}) 
can be written as \ $R(a_n) = \gamma\,,$ \ with \ $\gamma = 
\sqrt{2\hbar/(m\Omega)}\,\beta\,,$ \ and thus
\begin{equation}
\prod_{k=1}^{n}\left[\,\sum_{s=k}^{n}R(a_s)\right] = \prod_{k=1}^{n}\left[\,\gamma\,
(n-k+1)\right] = \gamma^n\,n!\,.
\label{eqfh1ho}
\end{equation}
On the other hand, the constant values of the potential parameters for (D) shape-invariant 
potential imply that for these systems we must have \ ${\cal  Z}_j = c\,,$ \ a constant. 
Using this and Eq.~(\ref{eqhnc}) we obtain
\begin{equation}
\prod_{k=0}^{n-1}{\cal  Z}_{j+k} = c^n\qquad\qquad\Longrightarrow\qquad\qquad
h_n\left(a_r\right) = {\sqrt{\gamma^n\,n!}\over c^n}\,.
\label{eqhnho}
\end{equation}
Taking into account it in Eqs.~(\ref{eqnorm}) and (\ref{eqzfn}) we find
\begin{equation}
{\cal N}\left(\vert z\vert^2;a_r\right) = \exp{\left(-{c^2\vert z\vert^2\over 2\gamma}\right)}\,,
\qquad\qquad {\rm and}\qquad\qquad
\vert z;a_r\rangle = e^{-\left(c\vert z\vert/\sqrt{2\gamma}\right)^2}
\sum_{n=0}^\infty {\left(cz/\sqrt{\gamma}\right)^n\over \sqrt{n!}}\,
\vert \Psi_n\rangle\,.
\label{eqzfnho}
\end{equation}
With these results we can show that the inner product (\ref{eqzpz}) of two 
coherent states can be readily found as
\begin{equation}
\langle z^\prime;a_r\vert z;a_r\rangle = \exp{\left[-{c^2\over 2\gamma}
\left(\vert z^\prime\vert^2
+\vert z\vert^2-2zz^{\prime *}\right)\right]}\,.
\label{eqzpzho}
\end{equation}
The overcompleteness property can be verified by using Eq.~(\ref{eqPhi}). The function
\begin{equation}
\Phi(\xi;a_r) = \sum_{n=0}^\infty \left({i\gamma\xi\over c^2}\right)^n = 
{1\over 1-i\gamma\xi/c^2}\,,
\label{eqPhiho}
\end{equation}
that has a pole at \ $\xi = -ic^2/\gamma$ \ and its integration in Eq.~(\ref{eqwrho1}) 
by using the lower-half complex plane enclosing this pole yields
\begin{equation}
{\cal W}(\rho;a_r) = {1\over 2\pi}\int_{-\infty}^\infty d\xi\,
{e^{-i\rho\xi}\over 1-i\gamma\xi/c^2} = e^{-c^2\rho/\gamma}\,.
\label{eqwrho2}
\end{equation}
Now, taking into account the result for ${\cal N}\left(\rho;a_r\right)$ and the relation between 
the function ${\cal W}(\rho;a_r)$ and the weight function, it is possible to show that \  
$w\left(\rho;a_r\right)= 1/\pi\,.$ \ 
The example of a (D) shape-invariant system is the harmonic oscillator
potential \ $V_-(x,a_{_1})\,,$ \ obtained with Eq.~(\ref{eqvpm}) and $W_{_{\rm D}}(x,a_{_1})$. 
In this case it should be noted that if we redefine the complex constant by \ 
$z\longrightarrow cz/\sqrt{\gamma}\,,$ \ and 
take into account that \ $\vert \Psi_n\rangle \rightarrow \vert n\rangle$ \ is 
an element of the Fock space \ 
$\left\{\vert n\rangle,\; n\ge0\right\}\,,$ \ 
we obtain for the coherent state and its inner product
\begin{equation}
\vert z;a_r\rangle = e^{-\vert z\vert^2/2}
\sum_{n=0}^\infty {z^n\over \sqrt{n!}}\,
\vert n\rangle\,,\qquad\qquad
\langle z^\prime;a_r\vert z;a_r\rangle = \exp{\left[-{1\over 2}\left(\vert z^\prime\vert^2
+\vert z\vert^2-2zz^{\prime *}\right)\right]}\,,
\label{eqzzin}
\end{equation}
which are the usual expressions for bosonic coherent states \cite{r3}. 
In this case we observe that the simplicity of the harmonic oscillator system 
does not permit any special modification in the standard result with the definition of 
the generalized coherent state (\ref{eqzfn}). 

For (C) shape-invariant systems, if we make the choice \ ${\cal  Z}_j = c\,,$ \ a constant, and 
following the steps above it is possible to obtain identical results as (D) shape-invariant 
systems for the coherent state. However, any other choice would imply different 
results. Just as an example, let us define the following auxiliary function 
\begin{equation}
g(a_j;c,d) = ca_j + d\,,
\label{eqfg}
\end{equation}
where $c$ and $d$ are constants. With the help of Eq.~(\ref{eqancd}) we can show that
\begin{equation}
\prod_{k=0}^{n-1}g(a_{j+k};c,d) =
{(-c\eta)^n\,\Gamma\left[n+j-\rho-d/(c\eta)-1\right]\over
\Gamma\left[j-\rho-d/(c\eta)-1\right]} = 
{(c\eta)^n\,\Gamma\left[\rho+d/(c\eta)-j+2\right]\over
\Gamma\left[\rho+d/(c\eta)-n-j+2\right]}\,,
\label{eqfrc}
\end{equation}
where \ $\eta = \sqrt{\hbar/(2m\Omega)}$ \ and \ $\rho = a_{_1}/\eta\,.$ \  
Taking into account this result and defining the functional \ ${\cal  Z}_j$ \ as
\begin{equation}
{\cal  Z}_j = \sqrt{g(a_{_1};-\gamma/\eta,1)}\;e^{-i\alpha R(a_{_1})}
\label{eqfgpa1}
\end{equation}
we get
\begin{equation}
\prod_{k=0}^{n-1}{\cal  Z}_{j+k} = 
\sqrt{\gamma^n\,\Gamma (n-\rho)\over\Gamma (-\rho)}\,e^{-i\alpha\gamma n}\,,
\label{eqfgpa}
\end{equation}
where we used that \ $e_n = n\gamma\,.$
Substituting Eqs.~(\ref{eqfh1ho}) and (\ref{eqfgpa}) in (\ref{eqhnc}) we obtain
\begin{equation}
h_n\left(a_r\right) = \sqrt{\Gamma (-\rho)\,\Gamma (n+1)\over \Gamma (n-\rho)}\,
e^{i\alpha\gamma n}\,,
\label{eqhncx}
\end{equation}
and we can show that the normalization factor (\ref{eqnorm}) in this case is given by
\begin{equation}
{\cal N}\left(\vert z\vert^2;a_r\right) = 
\left[{1\over \Gamma (-\rho)}\sum_{n=0}^\infty {\Gamma (n-\rho)
\over \Gamma (n+1)}\vert z\vert^{2n}\right]^{-1/2} = \left(1-\vert z\vert^2\right)^{-\rho/2}\,,
\label{eqnorc1}
\end{equation}
with the restriction \ $\vert z\vert<1\,.$ \ Thus, the coherent state (\ref{eqzfn}) obtained
with these results is
\begin{equation}
\vert z;a_r\rangle = \left(1-\vert z\vert^2\right)^{-\rho}\,
\sum_{n=0}^\infty \sqrt{\Gamma (n-\rho)\over \Gamma (-\rho)\,\Gamma (n+1)}\,
e^{-i\alpha\gamma n}\,z^n
\vert \Psi_n\rangle\,,
\label{eqzfnpt1}
\end{equation}
where we take \ $a_{_1}<0$ \ implying \ $\rho<0\,.$ As it is always possible 
to get \ $a_{_1}+\delta >0$ \ with an adequate choice of $\delta$, there are no problems 
with this assumption. In this case the inner product (\ref{eqzpz}) of two coherent states will be 
\begin{equation}
\langle z^\prime;a_r\vert z;a_r\rangle = 
\left[{\sqrt{\left(1-\vert z\vert^2\right)\,\left(1-\vert z^\prime\vert^2\right)}\over\,
\left(1-z^{\prime *}z\right)}\right]^{-\rho}\,.
\label{eqzpzc}
\end{equation}
The completeness can be obtained by using the measure \ $w\left(\vert z\vert^2;a_r\right) = 
-(\rho+1)\left(1-\vert z\vert^2\right)^{-2}/\pi\,,$ \
that is invariant on the disk $\vert z\vert<1\,.$ Example of a shape invariant type (C) 
system \cite{r24} is the double anharmonic potential \ $V_-(x,a_{_1})\,,$ \ obtained with 
Eq.~(\ref{eqvpm}) and using the superpotential \ $W_{_{\rm C}}(x,a_{_1})\,.$

The coherent state we obtained for the (C)-type shape-invariant system, Eq.~(\ref{eqzfnpt1}), is
the Perelomov coherent state \cite{r5b} for the group SU(1,1). This is not surprising since
the SU(1,1) algebra is both the shape-invariance and spectrum-generating algebra of this 
shape-invariant system. The appropriate realization of this algebra is 
\begin{equation}
\hat K_{_0} = {1\over 4}\left(\hat p^2+x^2+{\alpha\over x^2}\right)\qquad\quad {\rm and} 
\qquad\quad
\hat K_{_{\pm}} = {1\over 4}\left(\hat p^2-x^2+{\alpha\over x^2}\right)\pm 
{i\over 4}\left(\hat p x+x\hat p\right)\,.
\label{eqk0pm}
\end{equation}

For the (C)-type shape-invariant systems the shape-invariance \cite{r15} connects eigenstates of 
the same system.


\subsection{Types (A) and (B) shape-invariant systems}

The partner potentials $V_\pm(x)$ for these systems are obtained with
the superpotentials
\begin{eqnarray}
W_{_{\rm A}}(x,a_{_1}) &=& \sqrt{\hbar\Omega}\left\{\beta(a_{_1}+\gamma)
\cot{\left[\beta(x+\lambda)\right]}+\delta\csc{\left[\beta(x+\lambda)\right]}\right\}
\\
W_{_{\rm B}}(x,a_{_1}) &=& \sqrt{\hbar\Omega}\left[\beta(a_{_1}+\gamma)+
\delta\exp{\left(-\beta x\right)}\right]\,,
\label{eqwab}
\end{eqnarray}
being $\beta$, $\gamma$, $\delta$ and $\lambda$ real constants. For these
systems the remainders in the shape-invariant condition (\ref{eqsi}) are given by \ 
$R(a_{_1}) = \pm\beta^2\eta\,\left[2(a_{_1}+\gamma)\pm\eta\right]\,,$ \ 
with the potential parameters related by \ $a_{n+1} = a_n\pm\eta\,,$ \ 
where \ $\eta = \sqrt{\hbar/(2m\Omega)}$ \ and the signs $(+)$ and $(-)$ stand for (A) and (B) 
types, respectively. Using these results we can prove that for (A) type systems one has 
\begin{equation}
\prod_{k=1}^{n}\left[\,\sum_{s=k}^{n}R(a_s)\right] = 
{\kappa^{2n}\,\Gamma(n+1)\,\Gamma(2\rho+2n)\over\Gamma(2\rho+n)}\,,
\label{eqfhsab}
\end{equation}
with \ $\kappa = \eta\beta$ \ and \ $\rho = (a_{_1}+\gamma)/\eta\,.$
To investigate the consequences of our general approach for this type of systems let 
us consider some possibilities. First, if we make the choice
\ ${\cal  Z}_j = c\,,$ \ a constant, and use the result of Eqs.~(\ref{eqhnho}) and (\ref{eqfhsab})
we find 
\begin{equation}
h_n\left(a_r\right) = \sqrt{\Gamma(n+1)\,\Gamma(2\rho+2n)\over \Gamma(2\rho+n)}
\label{eqhsa}
\end{equation}
and
\begin{equation}
{\cal N}\left(\vert z\vert^2;a_r\right) = 
\left[\,\sum_{n=0}^\infty {\Gamma (2\rho+n)
\over \Gamma (n+1)\,\Gamma (2\rho+2n)}\vert z\vert^{2n}\right]^{-1/2}
\label{eqnorsa}
\end{equation}
for the expansion coefficient and the normalization factor, respectively, after assuming 
\ $c = \kappa\,.$ 
At this point, if we choose \ $\rho = 1/2$ \ we get the simple expression found
in \cite{r7} for the coherent state (\ref{eqzfn}) because in this case \ $h_n\left(a_r\right)
 = \sqrt{(2n)!}\,,$ \ 
and since by Eq. (\ref{eqnorm})
\begin{equation}
{\cal N}\left(\vert z\vert^2;a_r\right)^{-1} =
\sqrt{\sum_{n=0}^\infty {\vert z\vert^{2n}\over (2n)!}} = 
{1\over\sqrt{{\rm sech}\,{(\vert z\vert)}}}\,,
\qquad {\rm and} \qquad
\vert z;a_r\rangle = \sqrt{{\rm sech}\,{(\vert z\vert)}}\sum_{n=0}^\infty {z^n\over\sqrt{(2n)!}}\,
\vert \Psi_n\rangle\,.
\label{eqzsaf}
\end{equation}
As shown in \cite{r7}, in this case the identity resolution is obtained with the measure
\ $w\left(\vert z\vert^2;a_r\right) = e^{-\vert z\vert}/(2\vert z\vert)\,.$

Another interesting possibility is to use the auxiliary function (\ref{eqfg}). 
In this case, because of the translation relation between the $a_n$ potential parameters, 
we can prove that for type (A) systems one gets
\begin{equation}
\prod_{k=0}^{n-1}g(a_{j+k};c,d) =
{(c\eta)^n\,\Gamma\left[{\nu\over 2}+j+n+{d/(c\eta)}-1\right]\over
\Gamma\left[{\nu\over 2}+j+{d/(c\eta)}-1\right]}\,,
\label{eqfrsapt}
\end{equation}
where \ $\nu = 2a_{_1}/\eta\,.$ \ If we define the functional \ ${\cal  Z}_j = 
\sqrt{g\left(a_{_1};{2\kappa/\eta},\kappa\right)\;
g\left(a_{_1};{2\kappa/\eta},2\kappa\right)}\;e^{-i\alpha R(a_{_1})}\,,$ \ we obtain 
\begin{equation}
\prod_{k=0}^{n-1}{\cal  Z}_{j+k} = 
\sqrt{{(2\kappa)^{2n}\,\Gamma ({\nu\over 2}+n+1)\,\Gamma ({\nu\over 2} +n+{1\over 2})\over
\Gamma ({\nu\over 2}+1)\,\Gamma ({\nu\over 2}+{1\over 2})}}\,e^{-i\alpha e_n} = 
\sqrt{\kappa^{2n}\,\Gamma (\nu+2n+1)\over \Gamma (\nu+1)}\,e^{-i\alpha e_n}\,,
\label{eqfjksa}
\end{equation}
where \ $e_n = \kappa^2\,n(n+2\rho)\,.$
Assuming \ $\gamma=\eta/2$\ and using Eqs.~(\ref{eqfhsab}) and (\ref{eqfjksa}) in 
(\ref{eqhnc}) we obtain 
\begin{equation}
h_n\left(a_r\right) = \sqrt{\Gamma (\nu+1)\,\Gamma (n+1)\over \Gamma 
(\nu+n+1)}\,e^{i\alpha e_n}\,,
\label{eqhncta1}
\end{equation}
and we can show that the normalization factor (\ref{eqnorm}) in this case is given by
\begin{equation}
{\cal N}\left(\vert z\vert^2;a_r\right) = 
\left[{1\over \Gamma (\nu+1)}\sum_{n=0}^\infty {\Gamma (\nu+n+1)
\over \Gamma (n+1)}\vert z\vert^{2n}\right]^{-1/2} = \left(1-\vert z\vert^2\right)^{(\nu+1)/2}\,,
\label{eqnta1}
\end{equation}
with the restriction \ $\vert z\vert<1\,.$ \ The coherent state (\ref{eqzfn}) 
obtained with these results is
\begin{equation}
\vert z;a_r\rangle = \left(1-\vert z\vert^2\right)^{(\nu+1)/2}\,
\sum_{n=0}^\infty \sqrt{\Gamma (\nu+n+1)\over \Gamma 
(\nu+1)\,\Gamma (n+1)}\,e^{-i\alpha e_n}\,z^n
\vert \Psi_n\rangle\,.
\label{eqzfnfsa1}
\end{equation}
Comparing with \cite{r25} one notes that Eq.~(\ref{eqzfnfsa1}) is a form for the 
coherent state of the P\"oschl-Teller potential of first type \cite{r26}. This potential 
\ $V_-(x,a_{_1})\,,$ \ obtained with Eq.~(\ref{eqvpm}) and using the superpotential 
\ $W_{_{\rm A}}(x,a_{_1})\,,$ is the example of a shape-invariant system type (A). 
As show in \cite{r25}, in this case, the resolution of unity is obtained with the measure
\ $w\left(\vert z\vert^2;a_r\right) = \nu\left(1-\vert z\vert^2\right)^{-2}/\pi\,.$ \ 

Finally, note that if one takes
\begin{equation}
{\cal  Z}_j = {\sqrt{g\left(a_{_1};2/\eta,1\right)\,g\left(a_{_1};2/\eta,2\right)}
\over g\left[a_{_1};1/(\kappa\eta),(1+\nu/2)/\kappa\right]}\;e^{-i\alpha R(a_{_1})}\,,
\label{eqfgsa3}
\end{equation}
and follows the same way used before one get a second possible form for the coherent 
state of the P\"oschl-Teller potential \cite{r25,r27}
\begin{equation}
\vert z;a_r\rangle = {\vert z\vert^{\nu/2}\over\sqrt{I_{\nu}(2\vert z\vert)}}\,
\sum_{n=0}^\infty {e^{-i\alpha e_n}\,z^n\over \sqrt{\Gamma (n+1)\,\Gamma (\nu+n+1)}}\,
\vert \Psi_n\rangle\,,
\label{eqzfnfsa2}
\end{equation}
where \ $I_\nu(x)$ \ is the modified Bessel function of the first kind. As shown in 
Ref.~\cite{r25} and \cite{r27}, in this case the resolution of unity is given by the measure
\begin{equation}
w\left(\vert z\vert^2;a_r\right) = {2\over\pi}K_{\nu}(2\vert z\vert^2)\,,
\qquad{\rm with}\qquad
K_\nu(x) = {\pi\left[I_{-\nu}(x)-I_\nu(x)\right]\over 2\,\sin{(\pi\nu)}}\,,
\qquad \nu\not\in\Z \,.
\label{eqkmod}
\end{equation}

Note that the coherent state in Eq.~(\ref{eqzfnfsa2}) is the Barut-Girardello coherent state 
for the SU(1,1) algebra \cite{r5a}. This is not surprising since SU(1,1) is the 
shape-invariance algebra
for the P\"oschl-Teller potential as shown in Ref.~\cite{r15}. Note that in this case the 
shape-invariant potential relates a series of potentials with different depths, not the quantum
states of the given potential, i.e. the shape-invariance algebra is not the spectrum-generating
algebra in contrast to the (C) and (D) type shape-invariant systems. Hence the coherent 
state corresponds 
to a non-compact algebra with infinite number of states representing all possible potentials with 
different depths.

As a last example we obtain a new coherent state for this kind of systems with the
introduction of the functional
\begin{equation}
{\cal  Z}_j = \sqrt{g\left(a_{_1};\beta,\beta\gamma\right)\;
g\left(a_{_1};\beta,\beta\gamma+\kappa/2\right)\;
g\left(a_{_3};2/\eta,-\nu-2\sigma\right)\over
g\left(a_{_1};1/\eta,\rho+\gamma/\eta\right)\;
g\left(a_{_3};1/\eta,-\nu/2\right)}\;
e^{-i\alpha R(a_{_1})}\,,
\label{eqfgsa7}
\end{equation}
which leads to 
\begin{equation}
\prod_{k=0}^{n-1}{\cal  Z}_{j+k} = 
\sqrt{\kappa^{2n}\,\Gamma (2n+2\rho)\,\Gamma (n+2-\sigma)\over
\Gamma (2-\sigma)\,\Gamma (n+2\rho)\,\Gamma (n+2)}\,e^{-i\alpha e_n}\,.
\label{eqfjksa7}
\end{equation}
Therefore using Eqs.~(\ref{eqfhsab}) and (\ref{eqfjksa7}) in (\ref{eqhnc}) we obtain 
\begin{equation}
h_n\left(a_r\right) = \sqrt{\Gamma (2-\sigma)\,\Gamma (n+1)\,\Gamma (n+2)
\over \Gamma (n+2-\sigma)}\,e^{i\alpha e_n}\,,
\label{eqhncta7}
\end{equation}
and we can show that the normalization factor (\ref{eqnorm}) in this case is given by
\begin{equation}
{\cal N}\left(\vert z\vert^2;a_r\right) = 
\left[{1\over \Gamma (2-\sigma)}\sum_{n=0}^\infty {\Gamma (n+2-\sigma)
\over \Gamma (n+2)}{\vert z\vert^{2n}\over n!}\right]^{-1/2} = 
{1\over\sqrt{\Phi\left(2-\sigma;2;\vert z\vert^2\right)}}\,,
\label{eqnta7}
\end{equation}
where \ $\Phi(a;b;x) = _{_1}\!\!\!F_{_1}(a;b;x)$ \ is the degenerate hypergeometric function
\cite{r27a}. The coherent state of Eq.~(\ref{eqzfn}) obtained with these results is
\begin{equation}
\vert z;a_r\rangle = {1\over\sqrt{\Gamma (2-\sigma)\;
\Phi\left(2-\sigma;2;\vert z\vert^2\right)}}\,
\sum_{n=0}^\infty \sqrt{\Gamma (n+2-\sigma)\over \Gamma (n+2)\;
\Gamma (n+1)}\,e^{-i\alpha e_n}\,z^n\vert \Psi_n\rangle\,.
\label{eqzfnfsa7}
\end{equation}
With the help of the integral \cite{r27b}
\begin{equation}
\int_0^\infty t^{\lambda-1}\,e^{-t/2}\,{\rm W}_{\sigma,\mu}(t)\;dt = 
{\Gamma \left(\lambda-\mu-{1\over 2}\right)\;\Gamma \left(\lambda+\mu-{1\over 2}\right)
\over \Gamma \left(\lambda-\sigma+1\right)}\,,
\label{eqintg}
\end{equation}
it is possible to show that the resolution of the unity can be obtained with the measure
\begin{equation}
w\left(\vert z\vert^2;a_r\right) = {\Gamma (2-\sigma)\over \pi}\,e^{-(\vert z\vert^2/2)}\,
\Phi\left(2-\sigma;2;\vert z\vert^2\right)\,
{\rm W}_{\sigma,1/2}\left(\vert z\vert^2\right)\,,
\label{eqwkp7}
\end{equation}
where ${\rm W}_{\sigma,\mu}(x)$ is the Whittaker function \cite{r27a,r27b}.

One example of the shape-invariant systems of type (B) is the Morse potential \cite{r28}. 
This potential has a finite number of normalizable bound states which cannot form
a complete set of states in the Hilbert space, the condition necessary to construct the
coherent state using our generalized approach. We nevertheless observe that since the 
superpotential $W_{_{\rm B}}(x,a_{_1})$ has a special form ($x$-independent and linear in  
$a_{_1}$-term) 
it is possible construct coherent states for Morse potential systems using other sets 
of eigenstates that form a complete orthonormal basis in Hilbert space and
examples of these procedures can be found in the references \cite{r29,r30}. 

In the other hand, the shape-invariant systems classified as types (E) and (F) have 
superpotentials given by
\begin{eqnarray}
W_{_{\rm E}}(x,a_{_1}) = \sqrt{\hbar\Omega}\left(\beta a_{_1}\cot{\left[\beta(x+\lambda)\right]}+
{\delta\over a_{_1}}\right)\qquad {\rm and}\qquad
W_{_{\rm F}}(x,a_{_1}) = \sqrt{\hbar\Omega}\left({a_{_1}\over x}+{\delta\over a_{_1}}\right)\,,
\label{eqwef}
\end{eqnarray}
where $\beta$, $\delta$ and $\lambda$ are real constants. The systems classified as
type (E) only have bound states while the systems type (F) have continuous as well as bound 
states. 
Like the systems type (B), the (E) systems have a finite number of energy eigenstates. 
In this case, 
an alternative way is construct the coherent state using the finite set of energy eigenstates 
with
an adequate redefinition of the measure $w\left(\vert z\vert^2;a_r\right)$ to get a finite 
number of moments $\{\rho_n\}$, as was done in \cite{r31} for Morse potential. The systems of
type (F), i.e. the Coulomb potential, because of its three-dimensional character, present 
energy-degenerated eigenstates. In this case the expansion (\ref{eqzf1}) defined for our 
generalized coherent state must be appropriately adjusted for this situation. 
Our generalized definition (\ref{eqzf}) of coherent states for shape-invariant systems can be 
extended to include these alternative approaches for systems type (B), (E) and (F).  
More details and further developments on this subject will be published elsewhere.


\subsection{Self-similar potential systems}

All previous examples have partner potentials $V_\pm(x)$ with parameters related 
by a translation. One class of shape-invariant potentials are given by an infinite chain
of reflectionless potentials $V_\pm^{(k)}(x)\,,$ \ $(k=0,1,2,\dots)\,,$ 
\ for which associated superpotentials $W_k(x)$ satisfy the self-similar {\it ansatz}
\ $W_k(x) = q^k\,W(q^kx)\,,$ \ with \ $0<q<1\,.$ These set of partners potentials 
$V_\pm^{(k)}(x)$, also called self-similar potentials \cite{r32,r33}, have an infinite 
number of bound states and its parameters related by a scaling: \ 
$a_n = q^{n-1}a_{_1}\,.$ \ 
Shape invariance of self-similar potentials was studied in detail
in \cite{r34,r35}. In the simplest case studied by them the remainder of
Eq.~(\ref{eqsi}) is given by \ $R(a_{_1})= ca_{_1}\,,$ \ where $c$ 
is a constant. Hence 
\begin{equation}
\prod_{k=1}^{n}\left[\,\sum_{s=k}^{n}R(a_s)\right] = 
\left[{R(a_{_1})\over 1-q}\right]^n
q^{n(n-1)/2}\;(q;q)_n
\label{eqfh1ss}
\end{equation}
where the $q$-shifted factorial 
\ $(q;q)_n$ \ is defined as \ $(p;q)_0 = 1$ \ and \
$(p;q)_n = \Pi_{j=0}^{n-1}\,(1-pq^j)$, \ with $n\in\Z$. 
Coherent states for self-similar potentials were introduced in \cite{r7,r10,r11}.
Before to apply our generalized approach for this system let us first assume the choice
\ ${\cal  Z}_j = 1\,,$ \ and use it and the result of Eqs.~(\ref{eqfh1ss}) in the 
the expansion coefficient (\ref{eqhnc}) to show the coherent state (\ref{eqzfn}) 
in this case is given by
\begin{equation}
\vert z;a_r\rangle = {1\over\sqrt{E_q^{^{(-1/2)}}\left(\vert\xi_{_0}\vert^2\right)}}
\sum_{n=0}^\infty\,
{q^{-n^2/4}\over \sqrt{(q;q)_n}}
\,\xi_{_0}^n\,\vert \Psi_n\rangle\,.
\label{eqz4}
\end{equation}
where \ $\xi_{_0} = z\,\sqrt{(1-q)\biggl/\left[ \sqrt{q}R(a_{_1})\right]}$ \ and 
the $q$-exponential is defined by \cite{r36,r37,r38}
\begin{equation}
E_q^{^{(\mu)}}(x) = \sum_{n=0}^\infty {q^{\mu n^2}\over(q;q)_n}\,x^n\,.
\label{eqexpq}
\end{equation}
The result (\ref{eqz4}) is the normalized form of the initial expression we
obtained in our previous paper \cite{r10} for the coherent states of the 
self-similar potentials. To apply our generalized approach for this kind 
of potential system we assume
\begin{equation}
{\cal  Z}_j = R(a_{_1})\;e^{-i\alpha R(a_{_1})}\qquad {\rm yielding} \qquad
\prod_{k=0}^{n-1}{\cal Z}_{j+k} = [R(a_{_1})]^n\;q^{n(n-1)/2}\,e^{-i\alpha e_{_n}}\,,
\label{eqfgss}
\end{equation}
where \ $e_n = R(a_{_1})(1-q^n)/(1-q)\,.$ \
Substituting Eqs.~(\ref{eqfh1ss}) and (\ref{eqfgss}) in (\ref{eqhnc}) we find
\begin{equation}
h_n\left(a_r\right) = {\sqrt{(q;q)_n\over 
[R(a_{_1})\,(1-q)]^n\;q^{n(n-1)/2}}}\;e^{i\alpha e_n}\,,\quad\qquad {\rm and}\quad\qquad
{\cal N}\left(\vert z\vert^2;a_r\right) =
{1\over\sqrt{E_q^{^{(1/2)}}\left(\vert\xi_{_1}\vert^2\right)}}\,,
\label{eqnorss}
\end{equation}
where \ $\xi_{_1} = z\,\sqrt{R(a_{_1})(1-q)/\sqrt{q}}\,.$ 
The coherent state (\ref{eqzfn}) obtained with these results is
\begin{equation}
\vert z;a_r\rangle = {1\over\sqrt{E_q^{^{(1/2)}}\left(\vert\xi_{_1}\vert^2\right)}}
\sum_{n=0}^\infty\,
{q^{n^2/4}\over \sqrt{(q;q)_n}}\,e^{-i\alpha e_n}
\,\xi_{_1}^n\,\vert \Psi_n\rangle\,.
\label{eqz}
\end{equation}
In this case we can show that the inner product (\ref{eqzpz}) of two 
coherent states can be readily found as
\begin{equation}
\langle z^\prime;a_r\vert z;a_r\rangle = {E_q^{^{(1/2)}}\left(\xi_{_1}^{\prime *}
\xi_{_1}\right)\over\sqrt{ E_q^{^{(1/2)}}\left(\vert\xi_{_1}^\prime\vert^2\right)\,
E_q^{^{(1/2)}}\left(\vert\xi_{_1}\vert^2\right)}}\,.
\label{eqzpzss}
\end{equation}
This result still valid for the first expression obtained for coherent state of 
self-similar potentials (\ref{eqz4}) since we change \ $\xi_{_1}\rightarrow\xi_{_0}\,,$
\ and \ $E_q^{^{(1/2)}}(x)\rightarrow E_q^{^{(-1/2)}}(x)\,.$ \ 
Equation (\ref{eqz}) is the temporally stable version of the coherent state found in our
previous paper \cite{r11} (see also Ref.~\cite{r38a}). As shown in that paper, this 
choice for ${\cal Z}_j$ makes possible to establish an overcompleteness relation for 
the coherent state \ $\vert z;a_r\rangle$ \ using Ramanujan's integral extension 
of the beta function \cite{r39} and the measure, in this case, is given by
\begin{equation}
w\left(\vert z\vert^2;a_r\right) = {1\over
2\pi\;(-\vert\xi_{_1}\vert^2;q)_{_\infty}\,\log{(1/q)}}\,.
\label{eqwss}
\end{equation}

Finally we introduce a possible new coherent state for this 
class of shape invariant potentials by using the functional definition
\begin{equation}
{\cal  Z}_j = R(a_{_1})\sqrt{1-c/a_{_2}}\;e^{-i\alpha R(a_{_1})}\qquad\Longrightarrow\qquad
\prod_{k=0}^{n-1}{\cal Z}_{j+k} = [R(a_{_1})]^n\;q^{n(n-1)/2}\;
\sqrt{(c;q^{-1})_{n+1}\over 1-c}
\,e^{-i\alpha e_n}\,,
\label{eqfgssn}
\end{equation}
where $c$ is an arbitrary constant. 
Substituting Eqs.~(\ref{eqfh1ss}) and (\ref{eqfgssn}) in (\ref{eqhnc}) we find
\begin{equation}
h_n\left(a_r\right) = {\sqrt{(1-c)\;(q;q)_n\over 
[R(a_{_1})\,(1-q)]^n\;q^{n(n-1)/2}\;(c;q^{-1})_{n+1}}}\;e^{i\alpha e_n}\,,\qquad {\rm and}\qquad
{\cal N}\left(\vert z\vert^2;a_r\right) = 
\sqrt{(-c\vert\xi_{_2}\vert^2q^{-1};q)_{_\infty}\over (-\vert\xi_{_2}\vert^2;q)_{_\infty}}\,,
\label{eqnorssn}
\end{equation}
where \ $\xi_{_2} = z\,\sqrt{R(a_{_1})(1-q)}\,.$
The coherent state (\ref{eqzfn}) obtained with these results is
\begin{equation}
\vert z;a_r\rangle = \sqrt{(-c\vert\xi_{_2}\vert^2q^{-1};q)_{_\infty}\over 
(1-c)\;(-\vert\xi_{_2}\vert^2;q)_{_\infty}}\sum_{n=0}^\infty\,
q^{n(n-1)/4}\sqrt{(c;q^{-1})_{n+1}\over (q;q)_n}\,e^{-i\alpha e_n}
\,\xi_{_2}^n\,\vert \Psi_n\rangle\,.
\label{eqzssn1}
\end{equation}
In this case it is possible to establish an overcompleteness relation for the coherent state \
$\vert z;a_r\rangle$ \ with the introduction of the measure 
\begin{equation}
w\left(\vert \xi_{_2}\vert^2;a_r\right) = {R(a_{_1})(1-q)(1-c)\over
\pi q\,\log{(1/q)}}\left[{(-\vert\xi_{_2}\vert^2;q)_{_\infty}\over 
(-\vert\xi_{_2}\vert^2q^{-1};q)_{_\infty}}\right]\,,
\label{eqwssn}
\end{equation}
and using the Ramanujan integral given by \cite{r39}
\begin{equation}
\int_0^\infty t^{k-1}\,{(-at;q)_{_\infty}\over (-t;q)_{_\infty}}\,dt = 
{\log{(1/q)}\,(q;q)_{k-1}\over q^{k(k-1)/2}\,(a;q^{-1})_k}\,.
\label{eqintrama}
\end{equation}
(An elementary proof of (\ref{eqintrama}) was given by Askey \cite{r40}.)
Note that it is straightforward to show that the results for this last example reduce to the 
previous one when we take the limit \ $c\rightarrow 0$ \ and consider the properties of the
\ $(a;q)_{_\infty}$ \ functions and the relations between the $\xi_{_1}$ and $\xi_{_2}$ 
complex variables. Indeed the Ramanujan integral and its integral extension of the beta 
function \cite{r39} used in the previous example is a particular case of the more general 
Ramanujan integral (\ref{eqintrama}).  



\section{Final Remarks}

In this article, using an algebraic approach, we constructed generalized 
coherent states for shape-invariant systems. This generalization based on the
introduction of a factor functional ${\cal  Z}_j$ of the potential 
parameters in the coherent state: 
a) satisfies the set of essential requirements we enumerated in the Introduction 
to establish classical and quantum correspondence; b) reproduces results already 
known for shape-invariant potential systems;
c) gives new possible expressions for coherent states. Another aspect to emphasize is 
that our generalized construction of coherent states gives some insight on the question 
of relating different sets of coherent states found in the literature for such systems.

\section*{ACKNOWLEDGMENTS}

This  work   was supported in  part  by   the  U.S.  National  Science
Foundation Grants No.\ INT-0070889  and PHY-0244384 at the  University
of  Wisconsin, and  in  part by  the  University of Wisconsin Research
Committee   with  funds  granted by    the  Wisconsin Alumni  Research
Foundation. A.N.F.A. thanks to the Nuclear Theory Group at University 
of Wisconsin for their very kind hospitality.
 

\end{document}